# Axial localization of luminophores by partial coherence interferometry

Nicolas Sandeau, Hervé Rigneault, Pierre-François Lenne and Hugues Giovannini*
Institut Fresnel UMR CNRS 6133, Ecole Généraliste d'Ingénieurs de Marseille (EGIM), Université d'Aix-Marseille III, Domaine Universitaire Saint Jérôme, 13397 Marseille cedex 20, France

## ABSTRACT

We propose a solution for increasing the axial resolution of confocal microscopes. In the experimental set-up described in this paper an interference phenomenon between two counterpropagating beams is used to determine the axial position of a luminophore. The optical path difference between the two waves, which is related to the position of the luminophore, is recovered thanks to a second interferometer by using partial coherence interferometry demodulation technique. The proposed solution can find applications in biology for localizing with nanometric resolution a small number of tagged species.

## 1. INTRODUCTION

Strong efforts have been made in the last two decades for localizing luminescent species in particular for biology applications. For this purpose confocal microscopes have been developed and various detection schemes have been proposed for studying for example the properties of single molecules. The development of high numerical aperture microscope objectives has led to a strong lateral resolution (typically 300 nm for immersion objectives). However, due to the particular geometry of confocal microscopes, the axial resolution (1 micron is the typical value in this case) is not as good as the lateral one. This axial resolution is not sufficient to address problems related to the localization of tagged species in biology. Several solutions have been proposed to increase the resolution of confocal microscopes dedicated to the localization of luminescent targets. They roughly can be divided in two categories. In the first one the incident beam is structured thanks to adapted spatial filters[1]. The drawback in this case is that the incident beam presents ripples. In the second category, interference phenomena are used either by placing the sample in the vicinity of a mirror placed in front of the microscope objective[2] or by superimposing the two counterpropagating beams emitted by the luminophore placed between two microscope objectives. The second solution has led to the study and the development of the 4Pi microscope[3]. This solution has shown to give nanometric axial resolution and its field of potential applications is wide. However, as the Optical Path Difference (OPD) of the two interfering beams has to be maintained close to zero (smaller than the coherence length of the source) with very precise alignments of two arms of an interferometer whose lengths are greater than 10 cm, the 4Pi microscope, which requires accurate thermal and mechanical stabilization, is difficult to set-up.

In this paper we present a solution derivated from the 4Pi microscope in which the two interfering beams travel across almost the same path. In this case the OPD, which is related to the position of the luminophore between the two microscope objectives, is much greater than the coherence length of the source. The two interfering beams are sent to a interferometer and the OPD is determined by using Coherence Multiplexing[4] (also called Partial Coherence Interferometry (PCI)) technique. We recall the principles of the PCI, we describe the experimental set-up and discuss the effects of the luminophore displacement between the two microscope objectives on the output signal of the system.

* hugues.giovannini@fresnel.fr ; phone (33) 4 91 28 80 66 ; fax (33) 4 91 28 80 67

## 2. PARTIAL COHERENCE INTERFEROMETRY

Partial Coherence Interferometry (PCI) has been studied since 1962 by Mandel[5] and used in the early 80's in the field of telecommunications[6] or to encode the parameters to be measured by Optical Fiber Sensor systems[7]. In this technique a broadband light source illuminates an interferometer (that we will call SI, as "Sensing Interferometer"). In this case the power spectral density $S'(\sigma)$ of he light coming out from SI is given by (a visibility of the interference phenomenon equal to 1 is assumed) :

$$S'(\sigma) = S_0(\sigma).[1 + \cos(2\pi\sigma\Delta_s)],$$

where $\Delta_s$ is the OPD of SI, $\sigma$ is the wavenumber with $\sigma = 1/\lambda$ ($\lambda$ is the wavelength) and $S_0(\sigma)$ is the power spectral density of the light emitted by the source. We suppose that $\Delta_s$ is much greater than the coherence length $l_c$ of the source. As a consequence no interference can be observed when the light coming out from SI is recorded by a photodeterctor. However, the information concerning $\Delta_s$ is contained in the spectrum of the ouptut light. Indeed the period of modulation of $S'(\sigma)$ along $\sigma$ axis is equal to $\Delta_s$. This kind of modulated spectrum is often called "channeled spectrum". When S is sent (see Figure 1) to a second interferometer (that we will call DI, as "Demodulation Interferometer") whose OPD is $\Delta_d$, the power spectral density $S(\sigma)$ of the light coming out from DI is given by (here again a visibility of the interference phenomenon equal to 1 is assumed) :

$$S(\sigma) = S_0(\sigma).[1 + \cos(2\pi\sigma\Delta_s)].[1 + \cos(2\pi\sigma\Delta_d)].$$

When the output light is sent to a photodetector, the recorded signal is proportional to

$$R = \int_\sigma S(\sigma)d\sigma .$$

Figure 2 shows the typical shape of R as a function of $\Delta_d$ for $\Delta_d>0$ when $S_0(\sigma)$ is a Gaussian function.

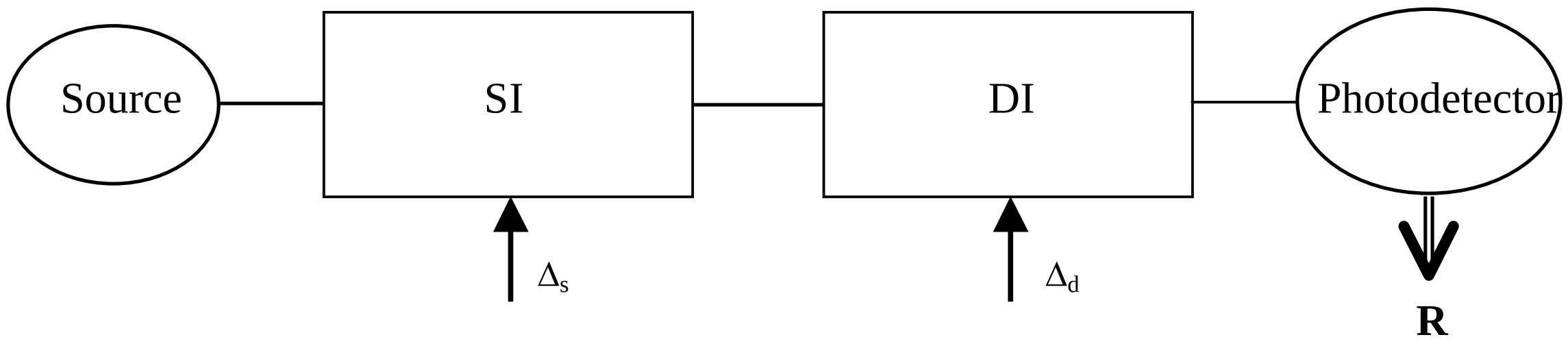

Fig. 1. PCI set-up. The two-interferometer system is illuminated by a broadband light source. The output signal recorded by the photodetector is R.

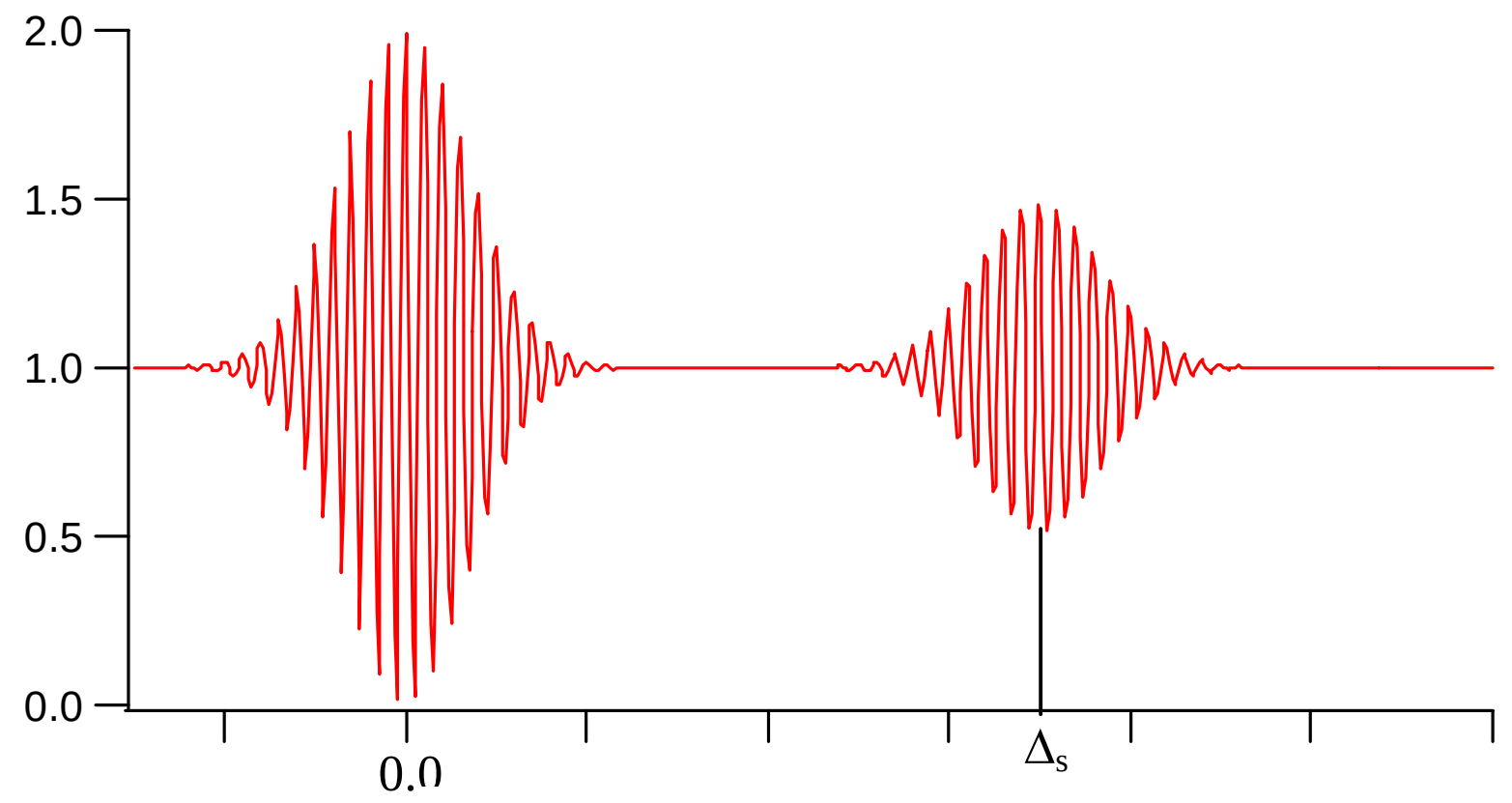

Fig. 2. Typical shape of signal R (in arbitrary unit) recorded by the detector as a function of the optical path difference of DI for $\Delta_s >> l_c$. In the calculation, $S_0(\sigma)$ in a Gaussian function.

The envelopes of the two peaks are proportional to the Fourier Transform of $S_0(\sigma)$. The first peak, which is the interferogram of the light source, is centered on 0 and does not contain any information about $\Delta_s$. The second peak corresponds to a correlation between the signals coming out from the two interferometers. When both $\Delta_s$ and $\Delta_d$ do not depend on $\sigma$, the maximum of this correlation peak is obtained when $\Delta_d=\Delta_s$ In this case the OPD of DI compensates that of SI and the interference phenomenon that occurs in SI is visible. One can see that any variation $\delta\Delta_s$ of $\Delta_s$ leads to the displacement of the second peak. The measurement of this displacement allows the determination of $\delta\Delta_s$. $\delta\Delta_s$ can be measured with accuracy[8] depending solely on the SNR ratio. Several solutions have been proposed and have shown that accuracy of typically better than 1 nm can be obtained.

## 3. SENSOR HEAD

We have designed the head of a modified confocal microscope in order to use PCI for localizing the position of a luminophore with nanometric resolution (see Figure 3).

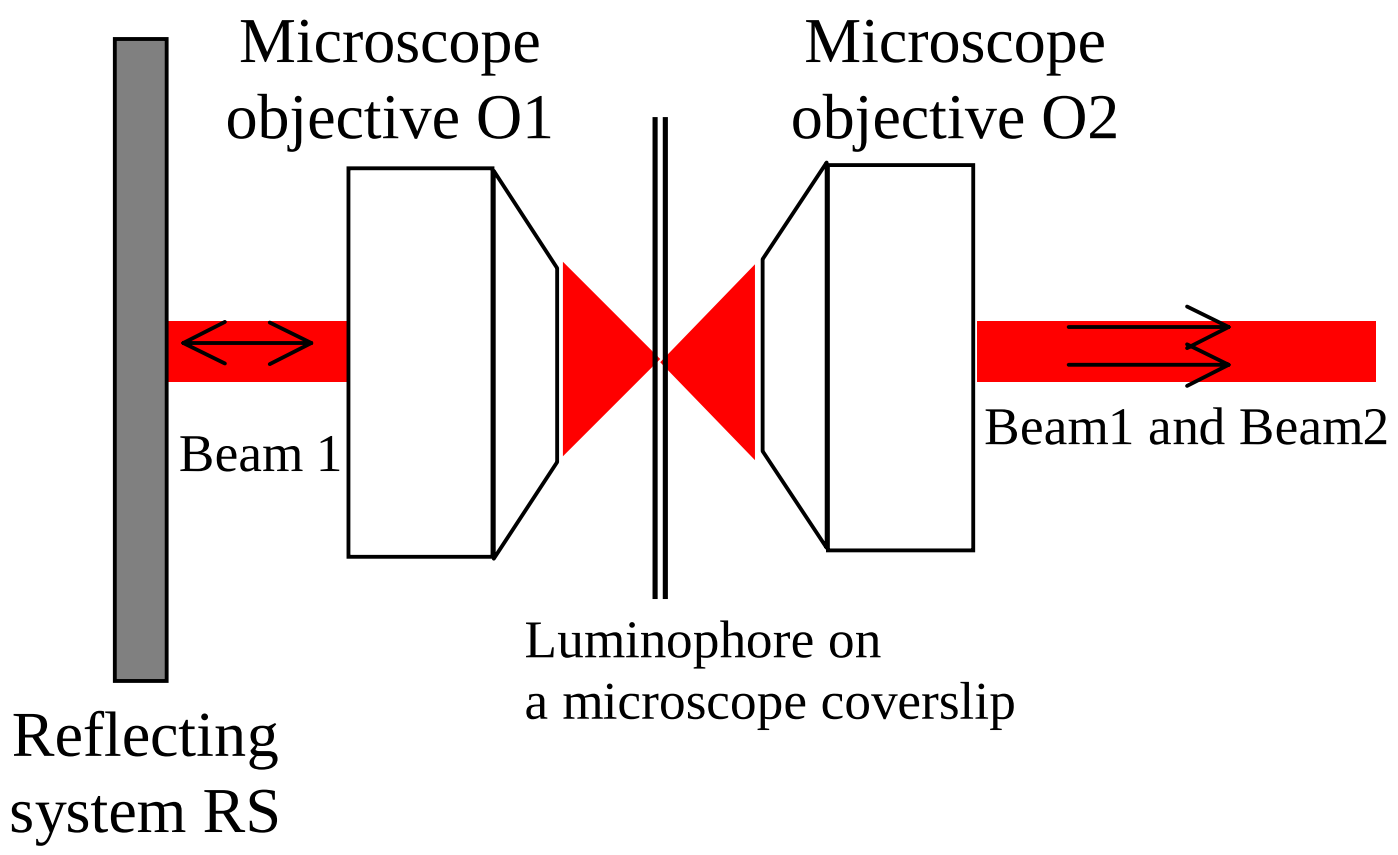

Fig. 3. Sensor head. Beam1 and Beam2 are parallel beams. The microscope objectives are assumed to be identical. They are emitted by a luminophore deposited on the surface of a microscope slide. They interfere and the OPD $\Delta_s$ depends on the position of the luminophore.

Here we assume that the luminophore is pumped by an external source. The luminophore is deposited on a microscope coverslip placed between two microscope objectives. The focuses of the two objectives are superimposed. We assume that the luminophore is at focus. Beam1, which is emitted on the left side of the luminophore (see Figure 3) is collimated by O1, reflected by RS, focused by O1 and collimated by O2. Beam2, which is emitted on the right side of the luminophore, is collimated by O2. Beam1 and Beam2 are parallel beams. They propagate in the same direction and interfere at infinity. The OPD, noted $\Delta_s$, between Beam1 and Beam2 depends on the position of the luminophore between O1 and O2. $\Delta_d$ is assumed to be much greater than the coherence length of the source (here the luminophore).

## 4. EXPERIMENTAL SET-UP

The light coming out from the sensor head is sent to a second interferometer whose OPD $\Delta_d$ can be varied. The luminophore is pumped by the laser beam. The laser light is rejected by the dichroïc mirrors. In Figure 4 we have represented the whole set-up when DI is a Michelson interferometer.

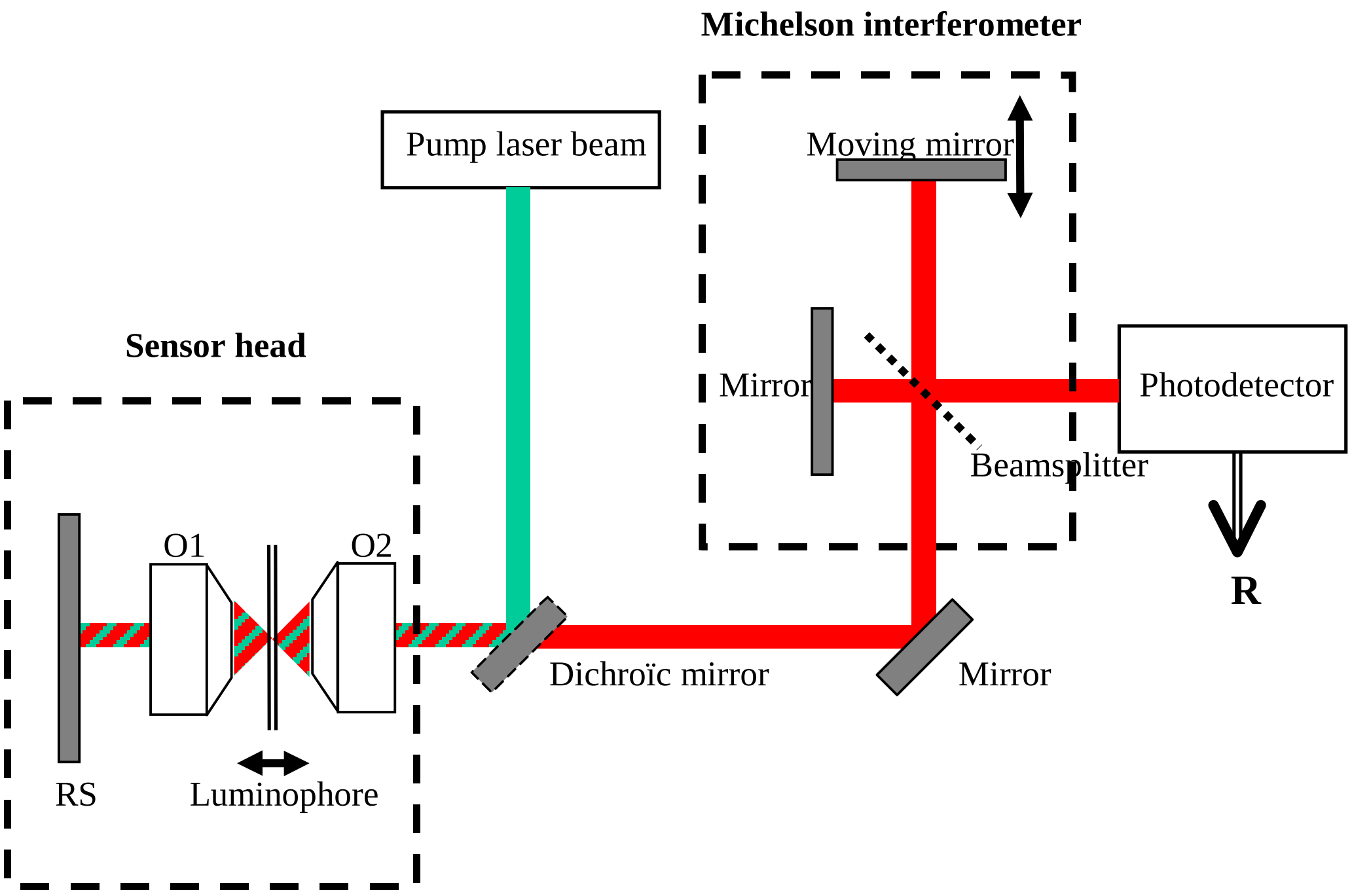

Fig. 4. Experimental set-up

The OPD $\Delta_d$ can be varied in order to record the correlation peak. The maximum value of the signal is obtained when $\Delta_s=\Delta_d$. By tracking the displacement of the correlation peak one can determine the variations of $\Delta_s$ with high accuracy.

## 5. DISCUSSION

The displacement of the luminophore between the two microscope objectives has several effects. Here we will only consider displacements along the optical axis of the sensor head. As said in section 3., these displacements lead to variations of $\Delta_s$ that can be recovered by analyzing the output signal R. Another effect is due to defocusing. When the luminophore is located in the common focal plane of the two microscope objectives, Beam1 and Beam2 are collimated. When the luminophore moves along the optical axis, the beams transmitted by the objectives are no more parallel and the visibility of the interference phenomenon is modified. This effect has widely been studied for analyzing the properties of Linnik microscopes[9]. It can be shown that the importance of this effect increases when the numerical aperture of the microscope objectives increases. Following the formula given in references 10 and 11, we have calculated the normalized output signal R for three positions of the luminophore along the optical axis of the system. The results are shown in Figure 5.

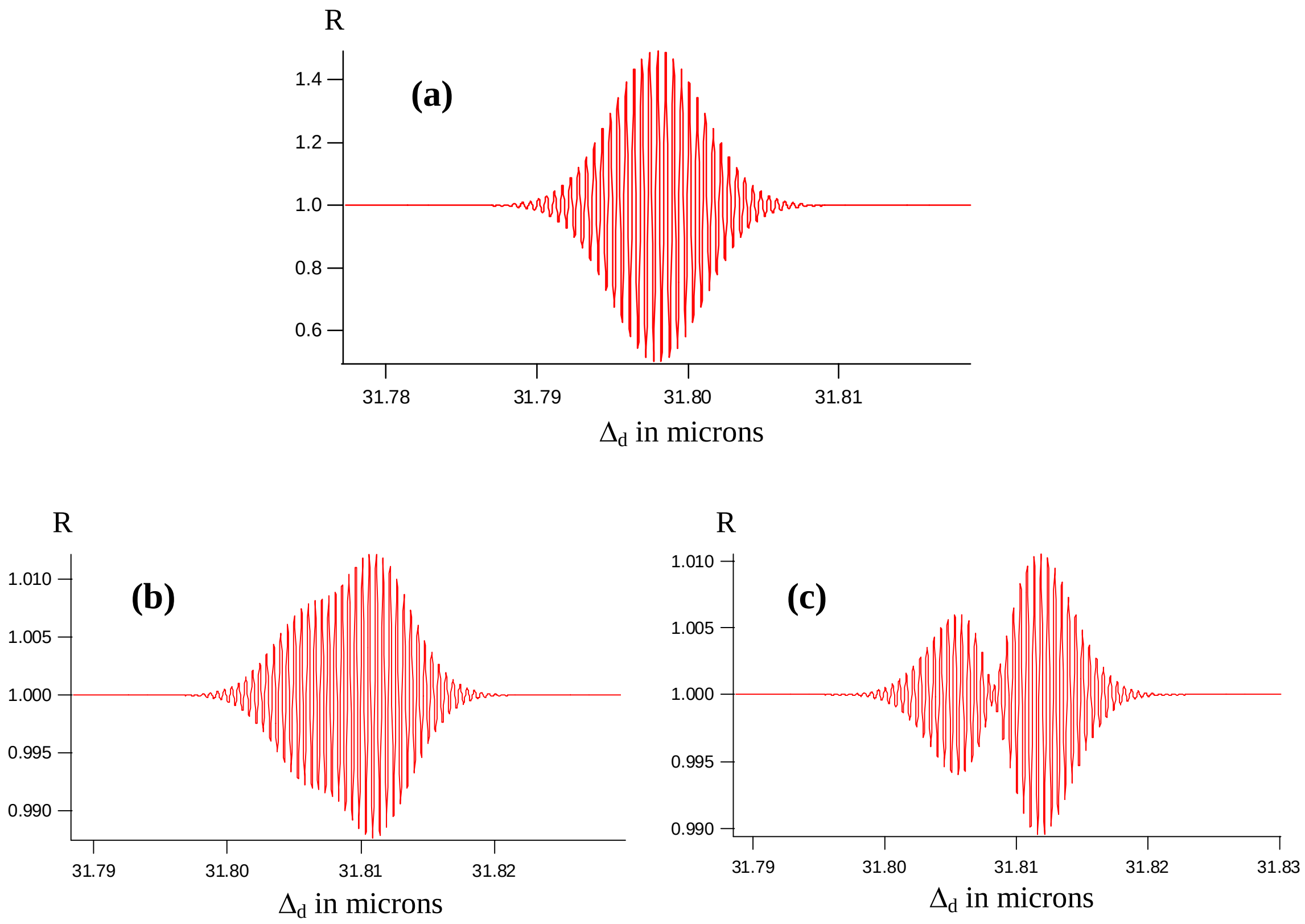

Fig. 5. Output signals signals R calculated for different positions of the luminophore along the optical axis of the system. The numerical aperture of the objective is equal to 0.3. The power spectral density of the light emitted by the luminophore is a Gaussian function centered at $\lambda_0$=525 nm with a FWHM=20 nm. (a) is calculated for the luminophore located at focus of the microscope objectives. (b) is calculated for a luminophore located at 2.8 µm from the focus. (c) is calculated for a luminophore located at 2.85 µm from the focus.

One can see from Figure 5 that, in our system, the displacement of the luminophore modifies the shape of the envelope of the correlation signal. The amplitude of the correlation peak decreases when the luminophore is displaced along the optical axis. Notice that the shape of the envelope of the correlation peak is not symmetrical. This property suggests a solution in order to determine if the luminophore has moved toward O1 or towards O2.

Another effect that must be taken into account is due to the interference fringes of the pump beam. There are two superimposed systems of fringes between the microscope objectives and the optical pump power received by the luminophore depends on its position. As the central wavelength of the light emitted by the laser and that emitted by the luminophore are different, the amplitude of the correlation signal is a product of two cosine functions.

Finally, to be more complete, one should study the effects of displacements perpendicular to the optical axis.

## 6. CONCLUSION

We have described an experimental set-up that should overcome certain difficulties due to the strong sensitivity to thermal and mechanical drifts in 4Pi microscopes. Indeed, here the two interfering beams propagate almost along the same optical path. This system should find applications for instance in biology for tracking tagged species or in metrology for localizing luminescent defects in transparent media.